\documentclass[final]{aipproc}
\usepackage{amsmath}
\usepackage{graphicx}
\layoutstyle{6x9}

\begin{document}

\title{Multipole analysis in  cosmic topology.}


\author{Peter Kramer}{
  address={Institut f\"ur Theoretische Physik der Universit\"at  T\"ubingen,\\ Germany}}

\begin{abstract}
Low multipole amplitudes in the Cosmic Microwave Background CMB radiation can be explained 
by selection rules from the underlying multiply-connected homotopy. 
We apply a multipole analysis to the harmonic bases and  introduce point symmetry.
We give explicit results for two cubic  3-spherical manifolds
and lowest polynomial degrees,  and derive three new spherical 3-manifolds.

\end{abstract}

\maketitle

\section{Introduction.}
\label{sec:intro}
Cosmology deals with the large-scale structure of the universe. A global description 
of this structure employs an averaging and smoothing of all the objects of astrophysics 
from solar systems to galaxies and clusters of galaxies. The result is a fluid model 
of the cosmos goverened by gravitation, as it was proposed by Einstein \cite{EI17}, see \cite{LAU56} pp. 160-164 and \cite{MI70}  pp. 704-710. This fluid model is governed by 
Einstein's differential equations which relate the space-time metric to the energy-momentum 
tensor. To device such a model one has to make assumptions about the global structure and topology of
the underlying space-time manifold. Einstein in his initial analysis of 1917 assumed  
cosmic 3-space in form of a 3-sphere. His assumption implies an average curvature $+1$.

The topology of cosmic 3-space has found new attention in relation to the Cosmic Microwave Background
(CMB) Radiation, which is supposed to originate from an early stage of the universe.
The fluctuations of the incoming CMB radiation are well described by a standard model, except for 
rather low amplitudes at the lowest multipole orders. This raised the question if multipole selection rules could be due to a particular topology of 3-space.  

\begin{figure}[t]
\includegraphics[width=1.0\textwidth]{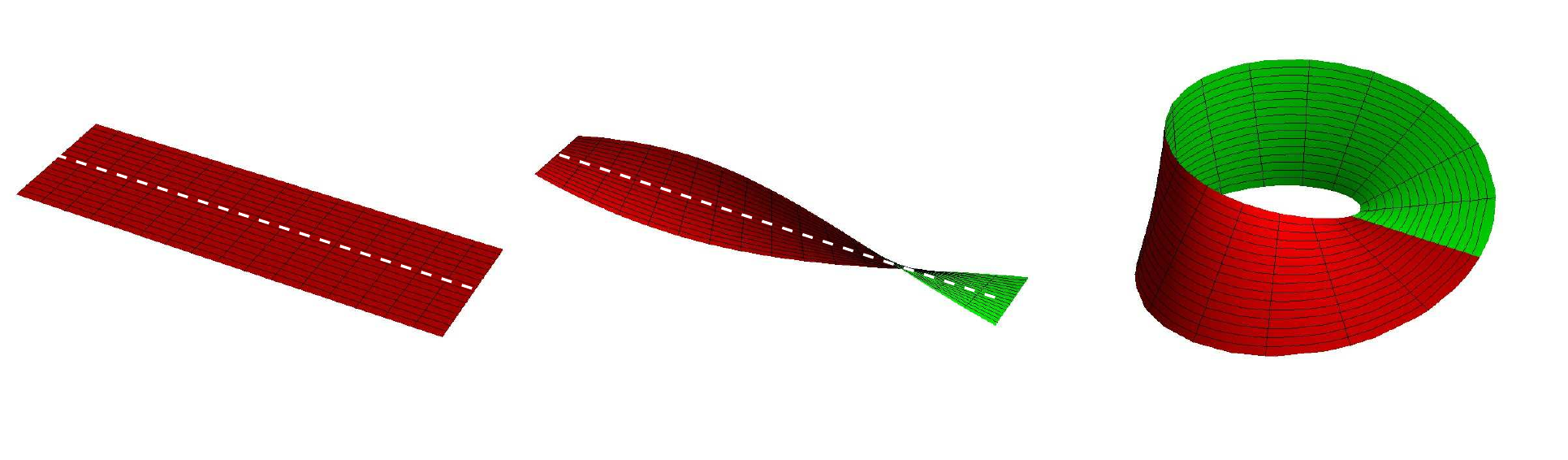}
\caption{\label{fig:moebi} {\bf Genesis of the M\"obius strip.} Left: as rectangular  cell of the M\"obius crystal {\bf cm} with broken glide reflection line, see Fig. \ref{fig:cm+8cellB},
Middle: Twisted by $\pi$  along the glide line,
Right: Bended into a circle and glued by homotopy.}
\end{figure}

\begin{figure}[t]
\includegraphics[width=0.9\textwidth]{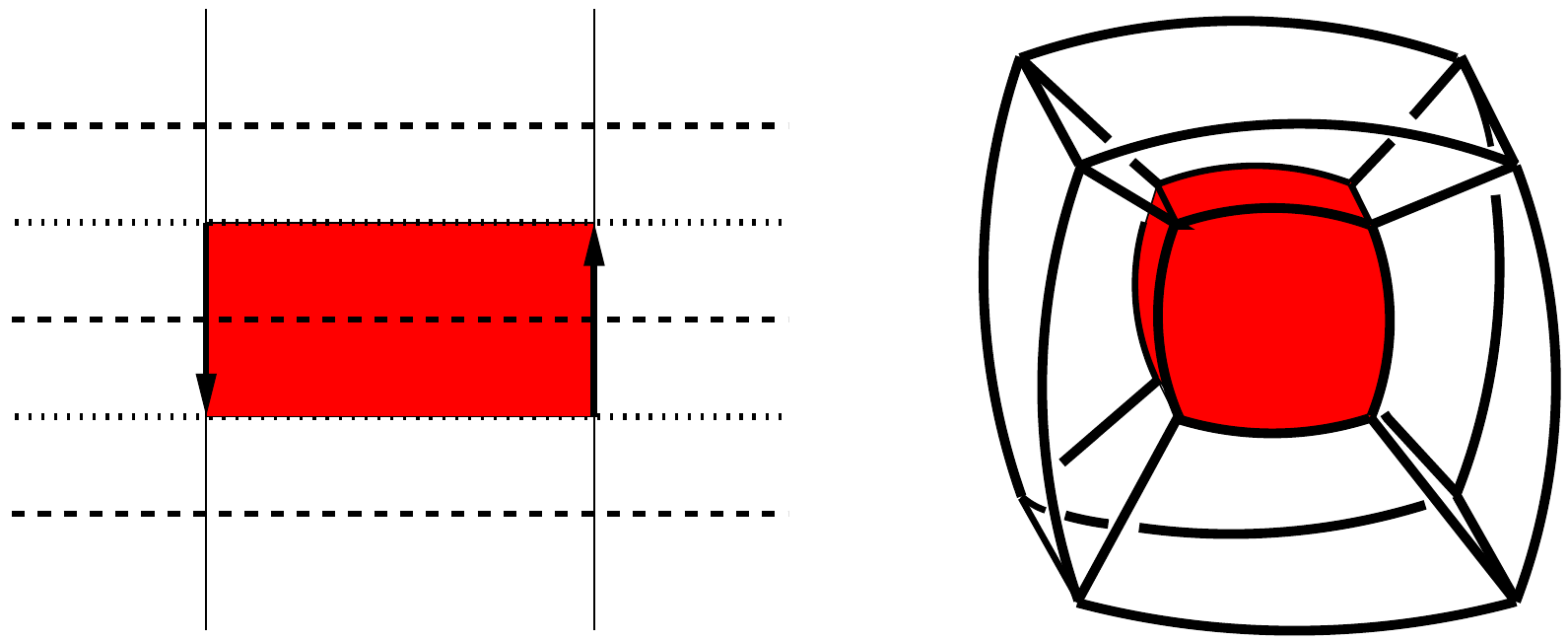}
\caption{\label{fig:cm+8cellB} {\bf M\"obius crystal and 8-cell.}
Left: The red M\"obius strip  tiles its cover, the plane, 
into the cells of a M\"obius crystal with 
crystallographic symbol {\bf cm}, with a broken glide line
and two pointed mirror reflection lines. The vertical edges are glued by homotopy, Fig.\ref{fig:moebi},
Right: A spherical red cube  tiles its cover, the 3-sphere, into the 8 spherical cubes of the 8-cell, shown in a projection.}
\end{figure}

A familiar paradigm for topology is the M\"obius strip. In \cite{KR10B} we explain its topological 
properties and relate it to the crystallographic space group {\bf cm}. A cell of this crystal 
is shown on the left-hand side of Fig. \ref{fig:cm+8cellB}. Any spherical topological 3-manifold
${\cal M}$  can be viewed as a prototile on Einstein's  3-sphere with pairs of faces glued according to the prescriptions of a group of homotopies. Among the spherical 3-manifolds are the Platonic ones.
Their homotopies were derived by Everitt \cite{EV04}. In \cite{KR10} and papers \cite{KR09},
\cite{KR09b} quoted therein we derived
from the homotopy or fundamental group of a Platonic manifold ${\cal M}$ the  corresponding groups 
deck(${\cal M}$) of deck operations acting on the 3-sphere. These groups generate by fixpoint-free action the tiling from the prototile. Each group of deck operations for a Platonic 3-manifold  ${\cal M}$ is 
constructed in \cite{KR10} as a subgroup of a Coxeter group $\Gamma$. Finally from the unimodular subgroups
of the Coxeter groups we construct three new spherical 3-manifolds.

\section{Actions and harmonic analysis on the 3-sphere.}
\label{sec:actions}

To study actions on the 3-sphere we pass from the set of four Cartesian coordinates $x=(x_0,x_1,x_2,x_3)$ in Euclidean 4-space to a matrix description,

\begin{equation}
\label{eq1}
 u=
\left[
\begin{array}{ll}
 z_1&z_2\\
-\overline{z}_2&\overline{z}_1\\
\end{array}
\right],\; z_1= x_0-ix_3,\: z_2=-x_2-ix_1,\: z_1\overline{z}_1+z_2\overline{z}_2=1.
\end{equation}
With the help of the Pauli matrices $\sigma_j$ and $\sigma_0=e$ and the trace $Tr$ we recover the 
Cartesian coordinates in the form
\begin{equation}
\label{eq2}
u=x_0\sigma_0-i\sum_{j=1}^3 x_j\sigma_j,\: 
x_0=\frac{1}{2} Tr(u\sigma_0),\: x_j=\frac{i}{2}\; Tr(u\sigma_j),\; j=1,2,3.
\end{equation}
The Wigner polynomials \cite{WI59}, \cite{ED57} are homogeneous polynomials $D^j_{m_1,m_2}(u)$ in the four complex variables $z_1,z_2,\overline{z}_1,\overline{z}_2$ of total degree $2j,\: j=0,\frac{1}{2},1,\frac{3}{2},...$, see \cite{KR10} Appendix A.
In the familiar Euler half-angles $(\alpha/2,\beta/2,\gamma/2)$, the Wigner polynomials take the form
\begin{eqnarray}
\label{eq4}
&&D^j_{m_1,m_2}(\alpha,\beta, \gamma )= \exp(im_1\alpha)d^j_{m_1,m_2}(\beta)\exp(im_2\gamma),
\\ \nonumber
&& -j\leq (m_1,m_2)\leq j. 
\end{eqnarray}
We look at the set of harmonic Wigner polynomials by starting from the integer or half-integer 
pairs of numbers $(m_1,m_2)$.
These  pairs of numbers  can be viewed as points from two nested lattices 
on a plane, see Fig.\ref{fig:gridN2N3}.
For given integer or half-integer values of $\{m_1,m_2\}$, one finds 
\begin{equation}
\label{eq5}
j=j_0 +\nu,\: \nu= 0,1,2,...,\: j_0={\rm Max}(|m_1|,|m_2|). 
\end{equation}
We say that any lattice point   $(m_1,m_2)$ in this plane  carries a {\em tower}, labelled by $\nu$, of Wigner polynomials $D^j$ according to eq. \ref{eq5}.
The Wigner polynomials form a complete orthogonal system of polynomial functions on the 3-sphere.
Moreover they are harmonic, that is, vanish under the application of the Laplacian acting 
on functions on Euclidean 4-space, see \cite{KR10}. Therefore they are a basis for harmonic analysis on the 3-sphere.

The isometric rotations of the group $SO(4,R) \sim (SU^l(2,C)\times SU^r(2,C)/K$
can be written in  the form $(g_l,g_r),\: g_l\in SU^l(2,C),\: g_r\in SU^r(2,C)$ and act on the coordinates 
$u$ as
\begin{equation}
\label{eq6}
 (g_l,g_r): u \rightarrow g_l^{-1}ug_r,\: K=\{(e,e),(-e,-e)\}. 
\end{equation}
The elements of the form  $(g,g),\: g\in SU^C(2,C) $ generate a subgroup $SU^C(2,C)$ acting on $u$ by conjugation.
The 3-sphere can be written as the homogeneous space $SO(4,R)/SU^C(2,C)$.
 We write the action eq. \ref{eq6} of $SO(4,R)$ on the Wigner polynomials as
\begin{eqnarray}
\label{eq6a}
&&(T_{(g_l,g_r)}D^j_{m_1,m_2})(u)=D^j_{m_1,m_2}(g_l^{-1}ug_r). 
\\ \nonumber 
&&= \sum_{m_1',m_2'} D^j_{m_1',m_2'}(u)D^j_{m_1,m_1'}(g_l^{-1})D^j_{m_2',m_2}(g_r).
\end{eqnarray}
Here we used the representation property of the Wigner polynomials.

A general pair $(g_l,g_r)$ can always be brought to diagonal form 
\begin{equation}
 \label{eq7}
g_l=c_l\delta_lc_l^{-1},\: g_r=c_r\delta_rc_r^{-1}. 
\end{equation}
We interprete the transformation 
\begin{equation}
 \label{eq8}
u \rightarrow u'=c_l^{-1}uc_r,
\end{equation} 
as a transformation to new coordinates $u'$. In these new coordinates,
$(\delta_l,\delta_r)$ are diagonal with diagonal entries $\delta_l:\exp(\pm i\alpha_l/2), 
\delta_r: \exp(\pm i\alpha_r/2)$, and the action eq. \ref{eq6a} with eq. \ref{eq7} takes the form
\begin{equation}
\label{eq9}
D^j_{m_1,m_2}(u) \rightarrow D^j_{m_1,m_2}((\delta_1)^{-1}u\delta_r)= \exp(i(-m_1\alpha_l+m_2\alpha_r))D^j_{m_1,m_2}(u). 
\end{equation}
Now we can go to a lattice description of the harmonic analysis for the two spherical cubic 
3-manifolds: The basis for the harmonic analysis consists of the towers of Wigner $D^j$ polynomials 
on top of a sublattice in the $(m_1,m_2)$-plane as given in Fig. \ref{fig:gridN2N3}.

\section{The spherical cubic manifolds.}
\label{sec:cubic}

As examples we shall choose the two spherical cubic manifolds $N2, N3$. The first homotopy groups 
of the Platonic spherical polyhedra were given in Everitt \cite{EV04}. In \cite{KR10} and work
cited therein we construct from Everitt's results the groups of deck transformations. These act on the 3-sphere and tile it into  Platonic polyhedra.

\section{The deck groups of the cubic  3-manifolds $N2, N3$.}
\label{subsec:N23}

\begin{figure}[t]
\includegraphics[width=0.5\textwidth]{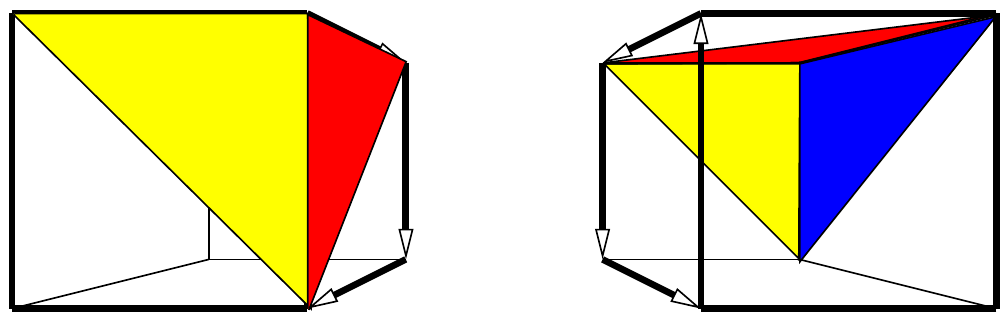} 
\caption{\label{fig:cubeglueB}
 Conversion of homotopic face gluing $F3\cup F1$ into a deck operation of two cubic tiles for the manifold N2.
The half-faces of $F1, F2, F3$ are marked by the colors yellow, blue, red respectively. For gluings of other faces see the left of Fig. \ref{fig:twocubetwist}}
\end{figure}

\begin{figure}[t]
\includegraphics[width=1.0\textwidth]{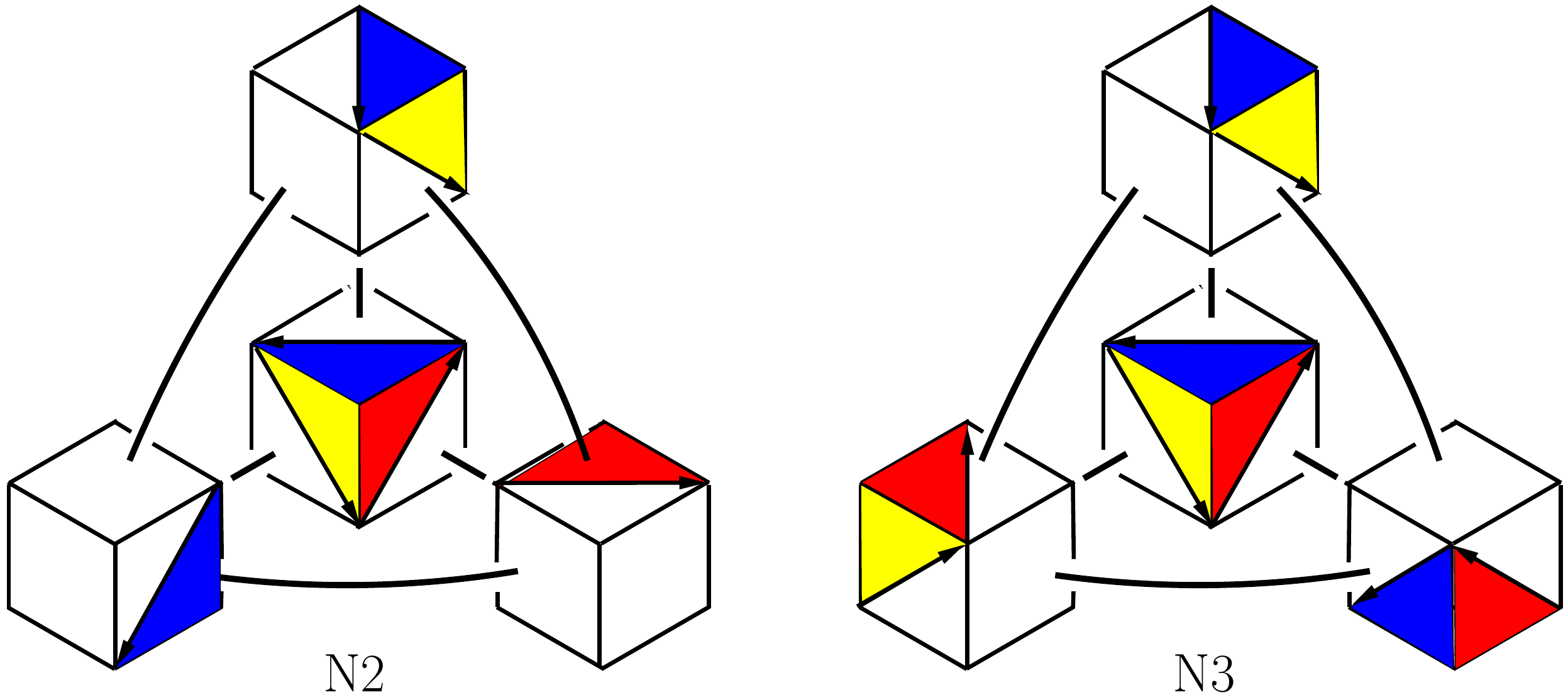} 
\caption{\label{fig:twocubetwist} {\bf The cubic manifolds $N2$ and $N3$}. The  cubic prototile and three neighbour tiles sharing 
its faces {\bf F1, F2, F3}. The four cubes are replaced by their  Euclidean counterparts  and separated from one another. Visible half-faces {\bf F1, F2, F3} are marked by the colors yellow, blue and red. The actions transforming the prototile into its three neighbours generate the deck transformations and the 8-cell tiling of $S^3$.}
\end{figure}

In  \cite{EV04} one  finds by an enumeration the homotopic face and edge gluings for this manifold.
The tiling of the 3-sphere is the 8-cell shown on the right in Fig \ref{fig:cm+8cellB}.
The cubic 3-manifold we take as the central spherical cube in the 8-cell.
The two different spherical cubic 3-manifolds differ in their face and edge gluing. 
In Fig. \ref{fig:twocubetwist} we have marked the faces with numbers $1, 2, 3$ by triangles 
with the colors yellow, blue, and red. The homotopic self-gluing of the initial 3-manifold is converted 
in the 8-cell tiling into a gluing of neighbour cubes sharing a face. This is illustrated in colors 
in Fig. \ref{fig:cubeglueB}. In Fig. \ref{fig:twocubetwist} we use the same color coding to
illustrate the two different next neighbour cubes for the two 3-manifolds $N2, N3$. 
 
\section{Harmonic analysis on the spherical cubic 3-manifolds $N2, N3$.}
\label{sec:harmonic}

Algebraically, the  deck operations, being rotations, contain an even number of Weyl reflections 
and can be written in terms of elements $(g_l,g_r) \in (SU^l(2,C)\times SU^r(2,C))$.

We now construct by projection the linear combinations of Wigner polynomials that span 
the harmonic analysis on the two cubic 3-manifolds.

\begin{figure}[t]
\includegraphics[width=0.8\textwidth]{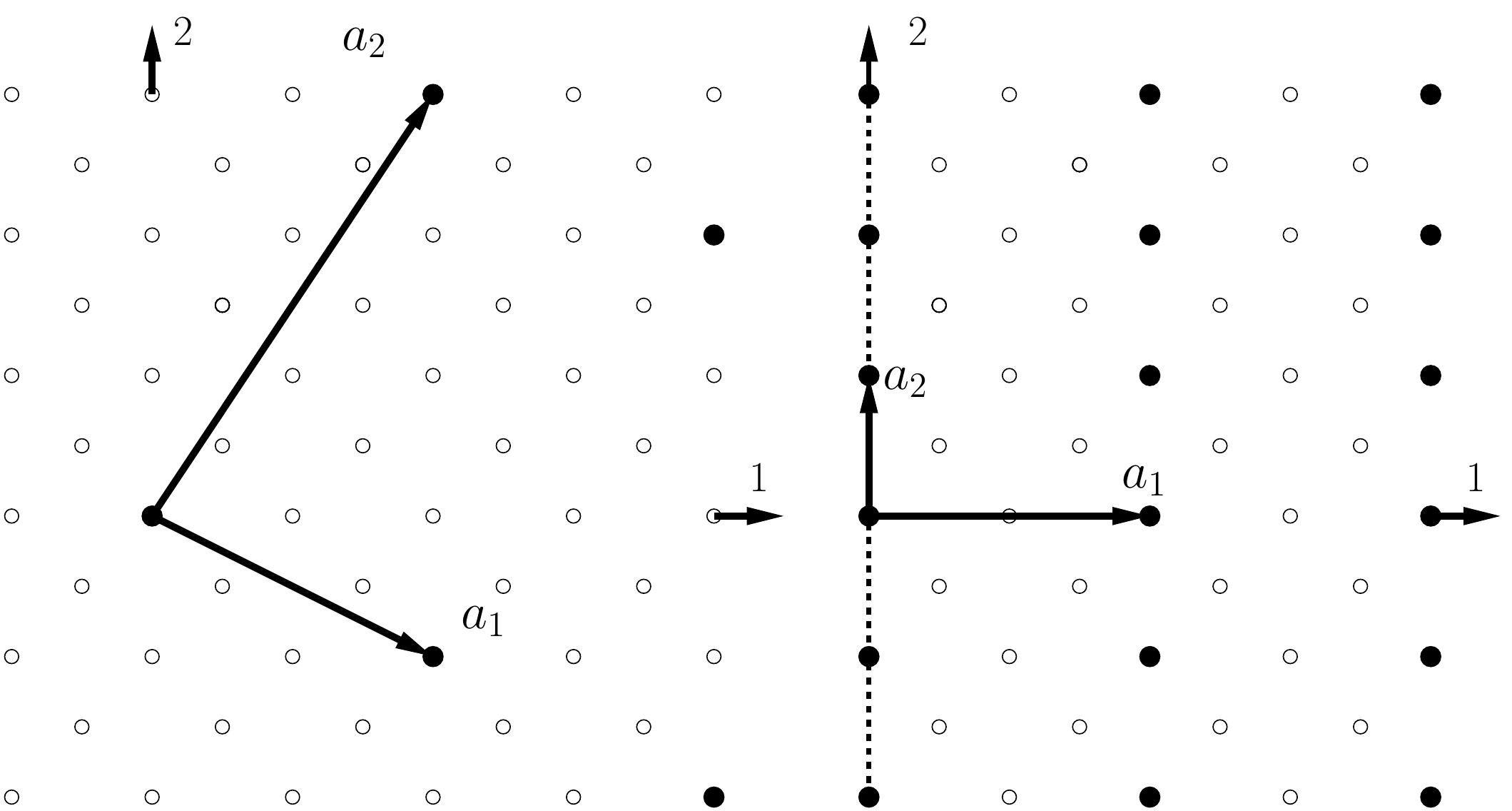} 
\caption{\label{fig:gridN2N3} {\bf Lattice representation of $N2$ and $N3$ basis:} Any lattice point $(m_1,m_2)$ carries the countable tower $D^j_{m_1,m_2}(u)$,
$j=j_0+\nu,\: \nu=0,1,2,...,\:j_0=Max(|m_1|,|m_2|)$ of Wigner polynomials.
Harmonic analysis on cubic  manifolds, left $N2$  , right $N3$ (with vertical mirror line), 
selects the  towers $D^j(u)$ on sublattices marked by  black points with sublattice bases $(a_1,a_2)$. 
Only these obey the homotopic boundary conditions.}
\end{figure}

For $N2$, the group $H={\rm deck}(N2)$ of deck transformations of the 8-cell tiling 
from \cite{KR10} is a cyclic group $C_8$. Its generator   $g_1=(g_l,g_r)$ is given in eq. \ref{eq11}.
We start from the set of Wigner polynomials and use their representation under $SO(4,R)$,
in terms of irreducible representations $D^j$ of $SU(2,C)$.
This allows to apply to them the projection operator to the identity representation 
of the group $H$,
\begin{equation}
\label{eq10}
(P^0D^j_{m_1m_2})(u)=  \sum_{m_1'm_2'} D^j_{m_1'm_2'}(u)\left[\frac{1}{|H|}\sum_{(g_l,g_r)\in H}\: D^j_{m_1m_1'}(g_l^{-1})D^j_{m_2'm_2}(g_r)\right]. 
\end{equation}
In general, eq. \ref{eq10} gives  a linear combination of Wigner polynomials. In case of the Platonic manifold $N2$, $H$  is a cyclic group. By a transformation $u \rightarrow u'$ of 
coordinates we can reduce the action of this cyclic group to diagonal form.
We start from the generator $g_1$ of the group ${\rm deck}(N2)=C_8$,
\begin{equation}
\label{eq11}
g_1=(g_l,g_r),\:
 g_l=\left[
\begin{array}{ll}
\overline{a}&0\\
0&a\\
 \end{array}
\right],\:
g_r=\left[
\begin{array}{ll}
0&\overline{a}\\
-a&0\\
 \end{array}
\right],\: a=\exp(i\frac{\pi}{4}).
\end{equation}
We diagonalize $g_r$ as 
\begin{equation}
\label{eq12}
 g_r= C \left[
\begin{array}{ll}
a^2& 0\\
0&-a^2\\
 \end{array}
\right] C^{-1},\:
C= \frac{1}{\sqrt{2}}
\left[
\begin{array}{ll}
-1&-a\\
\overline{a}&-1\\
 \end{array}
\right],\:
C^{-1}=C^{\dagger},\: {\rm det}(C)=1.
\end{equation}
Next we define new coordinates $u'$ by 
\begin{equation}
\label{eq13}
 u':=uC
\end{equation}
so that $g_1$ acts on $u'$ by diagonal matrices from left and right as
\begin{equation}
\label{eq14}
g_1:\: u' \rightarrow 
\left[
\begin{array}{ll}
a&0\\
0&\overline{a}\\
 \end{array}
\right] \: u'\:
\left[
\begin{array}{ll}
a^2&0\\
0&-a^2\\
 \end{array}
\right] =\delta_l^{-1}u'\delta_r.
\end{equation}
These relations allow to apply eq. \ref{eq9}.
It follows that we can reduce  the representation of $H$ to diagonal form.
Invariance under $H$ now gives a certain linear relation between $m_1$ and $m_2$.  
This relation singles out on the full lattice of points $(m_1,m_2)$ all the points of the 
sublattice with points
\begin{equation}
\label{eq15}
m_1+2m_2\equiv\: 0\: {\rm mod}\: 8.
\end{equation}
A basis of this sublattice is $a_1=(2,-1),\: a_2=(2,3)$, compare Fig. \ref{fig:gridN2N3}. 
The harmonic analysis on $N2$ is now given by the towers of Wigner polynomials 
on top of the black sublattice points.

For the cubic 3-manifold $N3$ we construct three glue generators $q_1, q_2, q_3$ in Table~\ref{TableN3a} from the homotopy group \cite{KR10}. The result for three faces is shown on the right of  Fig. \ref{fig:twocubetwist}.
The deck operations corresponding to the gluings generate  the quaternionic group $H=Q$, \cite{CO65} p. 134. It acts exclusively by left action on the 3-sphere.

\begin{table}
$\begin{array}{|l|l|l|l|} \hline
\label{c39}
i& q_i \: x&g_{li}&g_{ri}\\
\hline
1& (x_1,-x_0,x_3,-x_2)&
\left[
\begin{array}{ll}
0&-i\\
-i&0\\
\end{array}
\right]=-\mathbf{k}&e\\ \hline
2& (x_2,-x_3,-x_0,x_1)&
\left[
\begin{array}{ll}
0&-1\\
1&0\\
\end{array}
\right]=-\mathbf{j}&e\\ \hline
3& (x_3,x_2,-x_1,-x_0)&
\left[
\begin{array}{ll}
-i&0\\
0&i\\
\end{array}
\right]=-\mathbf{i}&e\\ \hline
\end{array}$
\caption{\label{TableN3a} 
The three generators $q_i=(g_{li}, g_{ri})$ of the quaternionic group $H={\rm deck}(N3)=Q$  as elements
of the Coxeter group $\Gamma$, and the corresponding  pairs 
$(g_{li},g_{ri})\in (SU^l(2,R)\times SU^r(2,R))$. Products of the matrices $\mathbf{i}$, 
$\mathbf{j}$, $\mathbf{k}$ follow the standard quaternionic rules.
}
\end{table}

For the harmonic analysis on $N3$ we employ the projection eq. \ref{eq10} for the quaternion group $Q$,
\begin{equation}
\label{eq16}
(P^0_{Q}D^j_{m_1,m_2})(u)= \frac{1}{8}\left[1+(-1)^{2j}\right] \left[1+(-1)^{m_1}\right] \left[D^j_{m_1,m_2}(u)+(-1)^jD^j_{-m1,m_2}(u)\right].
\end{equation}

\section{Harmonic analysis on Platonic cubic 3-manifolds.}
\label{sec:harmonicb}

In Fig. \ref{fig:gridN2N3}, we display the sublattices for the two cubic spherical  Platonic 3-manifolds. 
For the manifold $N3$ we put the symmetry eq. \ref{eq16}, $(m_1, m_2)\rightarrow (-m_1, m_2)$ as a vertical mirror line.
For the harmonic analysis 
it follows that an orthogonal  basis is given by the collection of all towers of Wigner polynomials on top of the sublattice points. Any basis function can be characterized by a sublattice point $(m_1,m_2)$ and by a number $\nu=0,1,2...$.

\section{Modelling incoming CMB by harmonic analysis.}
\label{sec:modelincom}

In this section we discuss  the algebraic tools for analysing incoming  CMB radiation 
in terms of the harmonic bases for a chosen topology.

\subsection{Alternative coordinates on $S^3$.}
\label{subsec:altern}
For the harmonic analysis on spherical 3-manifolds we use the spherical harmonics in the form of 
Wigner polynomials. These polynomials in the coordinates $x$ are often expressed in terms of 
Euler angle coordinates eq. \ref{eq4}.

An alternative system of polar coordinates is used by Aurich et al.~\cite{AU05}. Here 
\begin{equation}
\label{eq17}
\begin{array}{|l|} \hline
 x_0= \cos(\chi), \:x_1= \sin(\chi)\sin(\theta)\cos(\phi),
\\ 
 x_2=\sin(\chi)\sin(\theta)\sin(\phi),\: x_3= \sin(\chi)\cos(\theta),
\\ \hline
\\
 u=
\left[
\begin{array}{ll}
 \cos(\chi)-i\sin(\chi)\cos(\theta),& -i\sin(\chi)\sin(\theta)\exp(-i\phi)\\
 -i\sin(\chi)\sin(\theta)\exp(i\phi),    & \cos(\chi)+i\sin(\chi)\cos(\theta)
\end{array}
\right]\\ 
\\ \hline
\end{array}
\end{equation}
We shall see in eq.~\ref{eq21} that these polar coordinates are adapted to the analysis of incoming radiation in terms of its direction.

\subsection{Multipole expansion of  spherical harmonics on the 3-sphere.}
\label{subsec:multipole}

The CMB radiation as observed is given as a function of polar angles $(\theta, \phi)$ for its direction.
To compare with an expansion of the harmonic basis of a given 3-manifold, we must 
rewrite the Wigner polynomials in terms of polar angles. In terms of representation theory, this can be achieved by reducing the representations of $SO(4,R)$ into irreducible representations of its subgroup
$SU^C(2,C)$, see eq.\ref{eq19}. 

We relate our analysis algebraically to this description.

To adapt  the Wigner polynomials to a multipole expansion, we transform them 
for fixed degree $2j$ by use of Wigner coefficients of $SU(2,C)$, \cite{ED57} pp.~31-45, into the new harmonic polynomials
\begin{eqnarray}
\label{eq18}
 && \psi_{\beta l m}(u) = \delta_{\beta,2j+1}
\sum_{m_1,m_2}D^j_{m_1,m_2}(u) \langle j-m_1jm_2|lm\rangle (-1)^{j-m_1},
\\ \nonumber
&& l=0,1,..., 2j=\beta-1.
\end{eqnarray}

Whereas the index $j$ of the Wigner polynomials can be integer or half-integer, the multipole index $l$ takes only 
integer values. For fixed $l$ we have $2j\geq l$, and for fixed $2j$:  $0\leq l\leq 2j$.
Using representation theory of $SU(2,C)$ it can be shown from eq. \ref{eq6a}  that the conjugation action $u \rightarrow g^{-1}ug$ of the group $SU^C(2,C)$ acts by a rotation $R(g)$  only on the coordinate triple $(x_1,x_2,x_3)$, and  the new 
polynomials eq.~\ref{eq18} transform as 
\begin{equation}
\label{eq19}
(T_{(g,g)}\psi_{\beta l m})(u)=\psi_{\beta l m}(g^{-1}ug) = \sum_{m'=-l}^l\psi_{\beta l m'}(u)D^l_{m',m}(g),
\end{equation}
like the spherical harmonics $Y^l_m(\theta,\phi)$. We therefore adopt eq.~\ref{eq19} as the action of the usual rotation group for 
cosmological models covered by the 3-sphere, and  eq.~\ref{eq19} qualifies $l$ as the multipole index of incoming radiation.

The basis transformation eq.~\ref{eq18} is inverted with the help of the orthogonality of the Wigner coefficients \cite{ED57} to yield
\begin{equation}
\label{eq20}
 D^j_{m_1,m_2}(u)= \delta_{\beta, 2j+1} \delta_{m,-m_1+m_2}\sum_{l=0}^{2j}\psi_{\beta l m}(u) \langle j-m_1jm_2|lm\rangle (-1)^{j-m_1}.
\end{equation}
The result eq.~\ref{eq20} can be further elaborated  by  use of the alternative coordinates $(\chi, \theta, \phi)$  eq.~\ref{eq17}.
From \cite{AU05}, eqs. 9-17 we  find
 \begin{eqnarray}
\label{eq21}
 &&\psi_{\beta l m}(u)= R_{\beta l}(\chi)Y^l_m(\theta,\phi),\: \beta=2j+1,\: 2j\geq l,
\\ \nonumber
&&R_{\beta l}(\chi)=2^{l+\frac{1}{2}}l!\sqrt{\frac{\beta(\beta-l-1)!}{\pi(\beta+l)}}
C^{l+1}_{\beta-l-1}(\cos(\chi)).
\end{eqnarray}
where $C^{l+1}_{\beta-l-1}$ is a Gegenbauer polynomial. 
The alternative coordinates admit the separation of the new basis into a part depending on $\chi$ and a standard spherical harmonic as a function of polar 
coordinates $(\theta, \phi)$.

\begin{table}[t]
$
 \begin{array}{l|l}
l&Y^{\Gamma_1,l}=\sum_m a_{lm}Y^l_m(\theta, \phi)\\ \hline
0&Y^0_0\\
4&\sqrt{\frac{7}{12}}Y^4_0+\sqrt{\frac{5}{24}}(Y^4_4+Y^4_{-4})\\
6&\sqrt{\frac{1}{72}}Y^6_0-\sqrt{\frac{7}{144}}(Y^6_4+Y^6_{-4})\\
8&\frac{1}{64}\sqrt{33} Y^8_0+\frac{1}{12}\sqrt{\frac{21}{2}}(Y^8_4+Y^8_{-4})+
\frac{1}{24}\sqrt{\frac{195}{2}}(Y^8_8+Y^8_{-8})\\
\end{array}
$

\caption{\label{table:cubicl} The lowest cubic invariant  spherical harmonics $Y^{\Gamma_1,l}$, expressed by spherical harmonics $Y^l_m$.}
\end{table}

\begin{table}
$
 \begin{array}{l|l|l}
2j&l&\psi^{0,\Gamma_1,2j}=\sum_{l} b_{l} R_{2j+1\; l}(\chi)Y^{\Gamma_1,l}(\theta, \phi)\\ \hline
0&0& R_{1 0} Y^{\Gamma_1, 0}\\
4&0,4&\sqrt{\frac{2}{5}}R_{5 0}Y^{\Gamma_1, 0}+\sqrt{\frac{3}{5}}R_{54}Y^{\Gamma_1, 4}\\
6&0,4, 6&\sqrt{\frac{1}{7}}R_{7 0}Y^{\Gamma_1, 0}-
\sqrt{\frac{6}{11}}R_{7 4}Y^{\Gamma_1, 4}-\sqrt{\frac{24}{77}}R_{76}Y^{\Gamma_1, 6}\\
8&0,4, 6, 8&\frac{4}{3}\sqrt{\frac{1}{110}}R_{9 0}Y^{\Gamma_1, 0}
-\frac{12}{11}\sqrt{\frac{3}{65}}R_{9 4}Y^{\Gamma_1, 4}\\
&&+\frac{8\cdot 19}{165}R_{96}Y^{\Gamma_1, 6}
+\frac{4}{5}\sqrt{\frac{1}{33\cdot 13}}R_{98}Y^{\Gamma_1, 8}\\
 \end{array}
$
\caption{\label{table:cubicinv} The lowest $(Q \times_s O)$-invariant polynomials $\psi^{0,\Gamma_1,2j}$
of degree $2j$  on the 3-sphere in the basis, expressed by  the cubic invariant  spherical harmonics 
from  Table \ref{table:cubicl}. $(Q \times O)$-invariance enforces superpositions of several cubic invariant  spherical harmonics.}
\end{table}

\section{Point symmetry.}
\label{sec:points}

Any  Platonic spherical 3-manifold is distinguished by a specific  point symmetry group M  which 
stabilizes its center point. There arises the following enigma: The point group stabilizes the center point, but the deck group acts fixpoint-free. So the two groups can never  mix. Can they nevertheless 
be brought together, and what happens to the harmonic analysis? 

To examine this question we turn to the cubic spherical manifolds $N2, N3$. Their Coxeter group from \cite{KR10}  has the Coxeter diagram
\begin{equation}
 \label{eq21a}
\Gamma=\circ \stackrel{4}{-} \circ -\circ -\circ.
\end{equation}
The point group is the full cubic rotation group $O$ of order $|O|=24$. The group of deck transformations for
$N3$ is $H={\rm deck}(N3)=Q$, the quaternion group with eight elements. The cubic tiling of the 3-sphere 
is the 8-cell tiling of Einstein's 3-sphere, see \cite{SO58} p. 178. 
The relation between the deck and point groups is addressed in  Appendix C of  \cite{KR10}. 
In \cite{KR10B} one finds selection rules from point symmetry for the multipole orders $6\geq l\geq 0$.

For the cubic spherical manifold $N3$ we found there:

{\bf Prop 1}: The cubic point group 
$M=O$ of $N3$  under conjugation leaves invariant the group H=deck(N3)=Q, the quaternion group,  and with Q forms a semidirect group $Q \times_s O$, which turns out to be $S\Gamma$, the rotational
unimodular subgroup of the Coxeter group, generated by an even number of Weyl reflections.

The relation between point and deck group resembles the case of symmorphic space groups in 
Euclidean crystallography. There the commutative infinite translation group acts fixpoint-free, and a cubic cell has again the cubic group as its point group. The difference is that, when going from Euclidean 3-space to  the 3-sphere, the deck group is finite and no longer commutative.

If we consider first of all only the cubic point group $O$ as a subgroup of the rotation group $SO(3,R)$ in Euclidean 3-space,
there are well-known results from molecular physics for the multiplicity of its representation in a given representation of
$SO(3,R)$ with angular momentum $l$, see \cite{LA74} p. 438. For the identity representation denoted by $\Gamma_1$ of $O$, the lowest non-zero angular momentum is $l=4$. The cubic invariant linear combination of standard spherical harmonics for lowest values of $l$ are given in Table \ref{table:cubicl}.

Now we wish to include the quaternion group $H=Q$ of deck transformations. From the semidirect product property it follows that the projectors on the identity representation for $O$ and $Q$ commute with one another.
This allows for the following procedure: we take a cubic $O$-invariant linear combination of spherical harmonics and combine it according to eq. \ref{eq21} with the lowest possible function of the 
angle $\chi$. Then we transform this linear combination back by eq. \ref{eq20} into Wigner polynomials
and apply the projector eq. \ref{eq16} to   $Q$-invariant form. Next we transform back with eq. \ref{eq18} to the basis adapted to the multipole analysis. The resulting linear combination must still be 
$O$-invariant but may contain new $O$-invariant linear combinations of spherical harmonics. 
By use of  the cubic invariants from  Table \ref{table:cubicl} we obtain the fully $(Q \times_s O)$-invariant polynomials of 
Table \ref{table:cubicinv}. The construction requires only the Wigner coefficients of $SU(2,C)$ and
can easily be continued to higher polynomial degree. 

For the physics on the cubic spherical manifold with point symmetry,
there follows from Table \ref{table:cubicinv} a special and observable property: Different multipole orders
of spherical harmonics must be linearly combined to assure the overall invariance.

What happens with the first cubic spherical manifold $N2$ under cubic  point symmetry?
Here we have from \cite{KR10} the following universality: 

{\bf Prop 2}: Any particular homotopy of a regular polyhedron 
with fixed geometric shape implies a pairwise homotopic boundary condition on its faces. 
If full rotational symmetry is applied, all the faces and also all their edges are 
on the same footing. This implies that any particular homotopic boundary condition is automatically fulfilled. For the two cubic spherical manifolds $N2, N3$ it follows that the same rules apply
to their $S\Gamma$-invariant basis whose lowest part we give in Table \ref{table:cubicinv}.

The order of the semidirect product group $Q \times_s O$ is $|Q \times_s O|= 32\cdot 8=192$.
This is half the order 
$|\Gamma|=384$ of the Coxeter group. It means that we are projecting to the identity representations of $S\Gamma$.

\begin{figure}[t]
\includegraphics[width=0.5\textwidth]{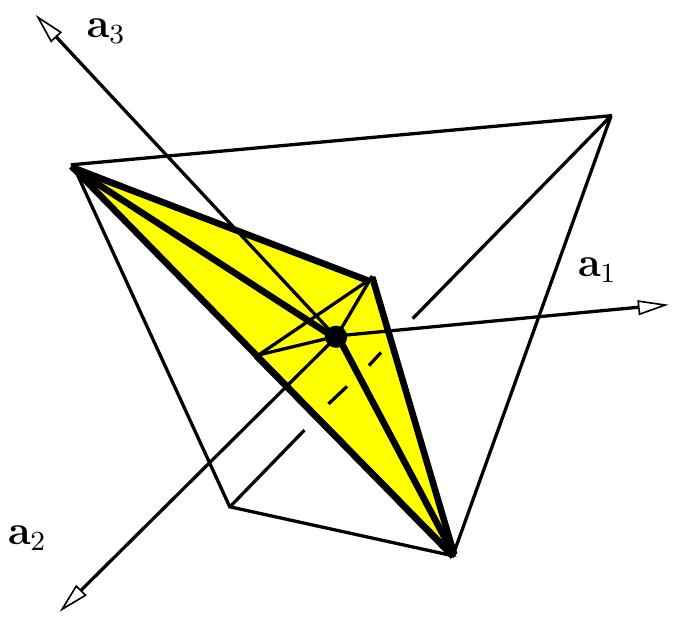} 
\caption{\label{fig:tetra1b} The new 3-manifold $N8$ as fundamental domain of the unimodular Coxeter group
S$\Gamma, \Gamma=\circ -\circ -\circ - \circ$ , glued from two Coxeter simplices.}
\end{figure}

\begin{figure}[t]
\includegraphics[width=0.6\textwidth]{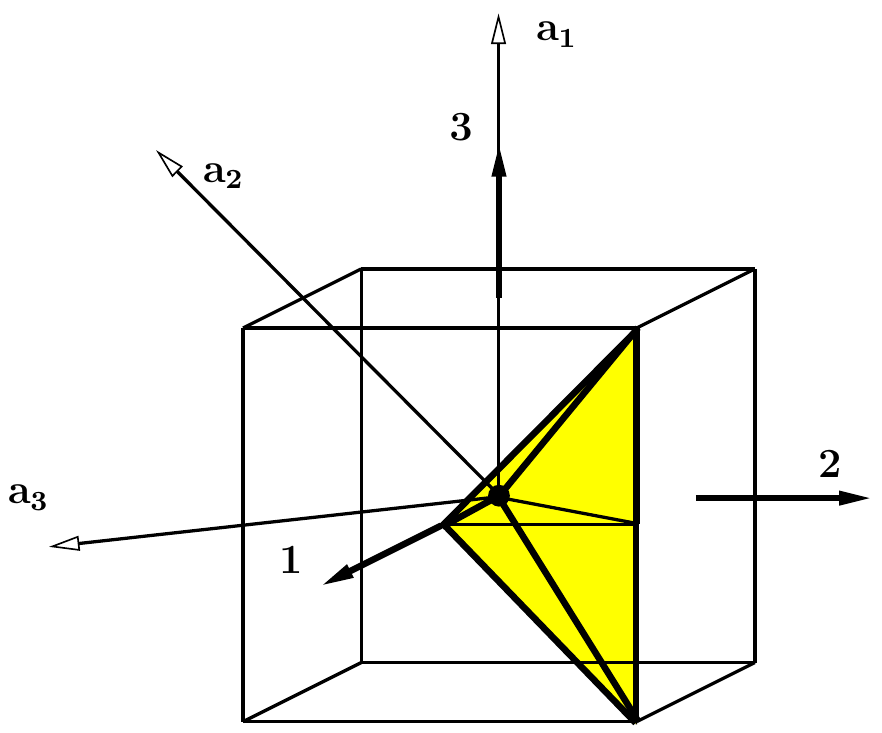} 
\caption{\label{fig:cubus1b} The new 3-manifold $N9$ as fundamental domain of the unimodular Coxeter group
S$\Gamma, \Gamma=\circ \stackrel{4}{-} \circ -\circ -\circ$ , glued from two Coxeter simplices.}
\end{figure}

\begin{figure}[t]
\includegraphics[width=0.6\textwidth]{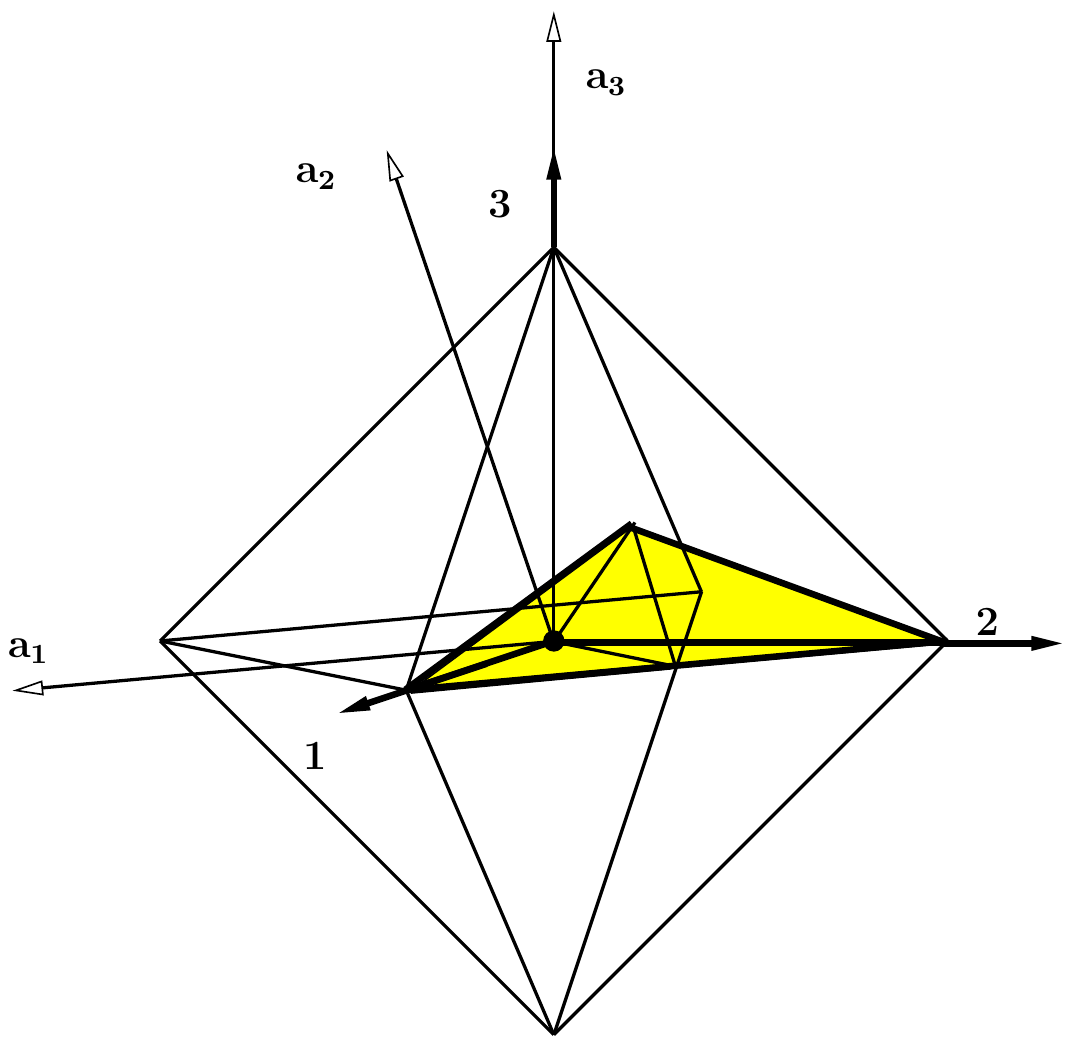} 
\caption{\label{fig:octa1b} The new 3-manifold $N10$ as fundamental domain of the unimodular Coxeter group
S$\Gamma, \Gamma=\circ -\circ \stackrel{4}{-}\circ - \circ$ , glued from two Coxeter simplices.}
\end{figure}

\section{New 3-manifolds from unimodular Coxeter groups.}
\label{sec:newmani}

We give a re-interpretation of the results of the previous section, 
We recall that in \cite{KR10} we constructed the spherical 3-manifolds from four Coxeter groups 
generated by Weyl reflections. Table \ref{table:Table2} gives the data. In the last section we found that under
inclusion of the cubic point group, the group ${\rm deck}(N3)$ extends into the unimodular 
subgroup S$\Gamma$ of the cubic Coxeter group $\Gamma$, generated by an even number of Weyl reflections.

 When we introduce in addition to topology the point symmetry of the spherical cube, we can define a fundamental subdomain on the cube  under the 
cubic point group. This subdomain may be taken as the cone, shown in Euclidean form in yellow on Fig. \ref{fig:cubus1b}. The cone is formed as a double simplex from two Coxeter simplices of the cubic Coxeter group $\Gamma$, with the second simplex the image of the first one under reflection in
the Weyl plane perpendicular to Weyl vector $a_1$. In the 8-cell tiling that covers the 3-sphere, the double simplex is a fundamental domain with respect to the unimodular Coxeter group S$\Gamma$, of volume fraction 
$1/(8\cdot 24)=1/192$.

This double simplex  on the 3-sphere forms a new  topological 3-manifold $N9$ with  the group  
${\rm deck(N9)}= S\Gamma, \Gamma=\circ \stackrel{4}{-} \circ -\circ -\circ $. With its small 
volume fraction it is an attractive candidate for cosmic topology. The first polynomials invariant under
${\rm deck}(N8)$ are the entries of Table \ref{table:cubicl}. 

Turning to the tretrahedral and octahedral 3-manifolds discussed in \cite{KR10}, their unimodular Coxeter groups admit  the analogeous construction. The unimodular subgroup for the tetrahedron is the even subgroup 
$A_5<S_5$. Its analysis in \cite{KR08} takes up  work with Marcos Moshinsky \cite{KR66}
on permutational symmetry.

We name the new 3-manifolds $N8,\: N9$, show their double simplices in Figs. \ref{fig:tetra1b}, \ref{fig:cubus1b}, and \ref{fig:octa1b},  and give their main data 
in Table \ref{table:Table1}, extended from Table 1 in \cite{KR10}. 
The double simplices are spherical counterparts to  the notion of asymmetric units as used in classical Euclidean 
crystallography. Since we know the geometry and the deck groups for  these new manifolds, it should not be hard to determine the corresponding homotopies. 
The harmonic analysis for ${\rm deck}(N8), {\rm deck}(N9)$  still has to be done. The harmonic bases will be
invariant in particular under the point group of the tetrahedron and octahedron respectively.
The lowest non-zero multipole index is $l=3$ for the tetrahedron and $l=4$ for the cubes.
If we wish to accomodate lower multipole order we must reduce the point symmetry of the manifold.
We conclude:

{\bf Prop 3}: The harmonic analysis on the three new 3-manifolds $N8, N9, N10$ strictly obeys the multipole selection rules given in \cite{KR10} Table 3 for the tetrahedral and cubic point groups respectively.

\begin{table}[t]
$
\begin{tabular}{|l|l|l|l|l|l|}\hline
Coxeter diagram $\Gamma$ & $|\Gamma|$ & Polyhedron ${\cal M}$ & $H={\rm deck}({\cal M})$ & $|H|$ & Reference \\ \hline
$\circ -\circ -\circ - \circ$               & $120$  & tetrahedron $N1$ & $C_5$         & $5$ & \cite{KR08} \\  \hline
$\circ \stackrel{4}{-} \circ -\circ -\circ$ & $384$  & cube $N2$        & $C_8$         & $8$ & \cite{KR09} \\                                                           &        & cube $N3$        & $Q$           & $8$ &\cite{KR09} \\  \hline
$\circ -\circ \stackrel{4}{-}\circ - \circ$ & $1152$ & octahedron $N4$  & $C_3\times Q$ & $24$ &\cite{KR09b} \\
                                            &        & octahedron $N5$  & $B$           & $24$ &\cite{KR09b} \\
                                            &        & octahedron $N6$  & ${\cal T}^*$  & $24$&\cite{KR09b} \\ \hline
$\circ -\circ -\circ \stackrel{5}{-} \circ$ & $120\cdot 120$ & dodecahedron $N1'$ & ${\cal J}^*$  & $120$ & \cite{KR05}, \cite{KR06} \\ \hline
$\circ -\circ -\circ - \circ$&$120$&double\: simplex\;$N8$&S$\Gamma$&$60$&\\ \hline
$\circ \stackrel{4}{-} \circ -\circ -\circ $&$384$&double\:simplex\;$N9$&S$\Gamma$&$192$&\\ \hline
$ \circ -\circ \stackrel{4}{-}\circ - \circ$&$1152$&double\:simplex\;$N10$&S$\Gamma$&$576$&\\
\hline
\end{tabular}
$
\caption{\label{table:Table1}
4 Coxeter groups $\Gamma$, 4 Platonic polyhedra ${\cal M}$, 10 groups $H={\rm deck}({\cal M})$ of order $|H|$.
$C_n$ denotes a cyclic, $Q$ the quaternion, ${\cal T}^*$ the binary tetrahedral, ${\cal J}^*$ the binary icosahedral, S$\Gamma$ a unimodular Coxeter group. The symbols $Ni$ are generalized from  \cite{EV04}.}
\end{table}

\begin{table}[t]
$
\begin{array}{|l|l|l|l|l|} \hline
\Gamma & a_1 & a_2 &a_3& a_4\\ \hline
\circ -\circ -\circ - \circ &  (0,0,0,1)&(0,0,\sqrt{\frac{3}{4}},\frac{1}{2})
& (0,\sqrt{\frac{2}{3}},\sqrt{\frac{1}{3}},0)& (\sqrt{\frac{5}{8}},\sqrt{\frac{3}{8}},0,0)\\ \hline
\circ \stackrel{4}{-} \circ -\circ -\circ&  (0,0,0,1)& (0,0,-\frac{1}{\sqrt{2}},\frac{1}{\sqrt{2}})
&(0,\frac{1}{\sqrt{2}},-\frac{1}{\sqrt{2}},0)&(-\frac{1}{\sqrt{2}},\frac{1}{\sqrt{2}},0,0)\\ \hline
\circ -\circ \stackrel{4}{-}\circ - \circ& (0,\frac{1}{\sqrt{2}},-\frac{1}{\sqrt{2}},0)&(0,0,-\frac{1}{\sqrt{2}},\frac{1}{\sqrt{2}})
&(0,0,0,1)& (\frac{1}{2},\frac{1}{2},\frac{1}{2},\frac{1}{2})\\ \hline
\circ -\circ -\circ \stackrel{5}{-} \circ& (0,0,1,0)&(0,-\frac{\sqrt{-\tau+3}}{2},\frac{\tau}{2},0)
&(0,-\sqrt{\frac{\tau+2}{5}},0,-\sqrt{\frac{-\tau+3}{5}})&(\frac{\sqrt{2-\tau}}{2},0,0,-\frac{\sqrt{\tau+2}}{2}) \\ \hline
\end{array}
$
\caption{\label{table:Table2}
The Weyl vectors $a_s$  
for the four Coxeter groups $\Gamma$ from Table~\ref{table:Table1} with  $\tau:=\frac{1+\sqrt{5}}{2}$.
}
\end{table}

\section{Conclusion.}
\label{sec:concl}

On the example of the cubic spherical 3-manifolds, we have explained  the construction of
the harmonic analysis from topology and its transformation into an expansion for the CMB radiation,
ordered by the multipole index $l$. We implemented 
the additional assumption of point symmetry for spherical manifolds. This assumption yields 
strong selection rules, including a lowest non-trivial multipole order $l$. Similar rules apply to the other Platonic spherical manifolds analyzed in \cite{KR10}. These strong selection rules are easier
to test from the fluctuation spectrum of the CMB radiation. Moreover we have shown that the inclusion 
of the unimodular Coxeter groups S$\Gamma$ yields $3$ new topological 3-manifolds which cover rather small
fractions of the 3-sphere.
  
\section{Acknowledgment.}
\label{sec:acknow}
This meeting is devoted to the memory of our great teacher and good friend, Professor Marcos Moshinsky. For over five decades I had the chance to share with him  his insight into groups,  their representations and applications in physics. The initial steps of the present analysis were  discussed with him in 2008 in Mexico. We all miss Marcos, and we shall never forget him.

\end{document}